\documentstyle[preprint,psfig,aps,prl]{revtex}
\tighten
\begin{document}

\draft

\preprint{\today}
\title{A Lattice--Boltzmann method for the simulation of transport 
       phenomena in charged colloids}
\author{J\"urgen Horbach \footnote{Author to whom correspondence
        should be addressed. e-mail: horbach@amolf.nl} and Daan Frenkel}
\address{Kruislaan 407, 1098 SJ Amsterdam, Netherlands}
\maketitle

\begin{abstract}
We present a new simulation scheme based on the Lattice--Boltzmann
method to simulate the dynamics of charged colloids in an electrolyte.
In our model we describe the electrostatics on the level of a
Poisson--Boltzmann equation and the hydrodynamics of the fluid by the
linearized Navier--Stokes equations. We verify our simulation scheme
by means of a Chapman--Enskog expansion. Our method is applied to the
calculation of the reduced sedimentation velocity $U/U_0$ for a cubic
array of charged spheres in an electrolyte.  We show that we recover the
analytical solution first derived by Booth~(F. Booth, J. Chem. 
Phys. {\bf 22}, 1956 (1954)) for a weakly
charged, isolated sphere in an unbounded electrolyte.  The present method
makes it possible to go beyond the Booth theory, and we discuss the
dependence of the sedimentation velocity on the charge of the spheres.
Finally we compare our results to experimental data.
\end{abstract}
\pacs{PACS numbers: 47.65.+a, 82.70.Dd, 47.11.+j, 83.85.Pt, 05.20.Dd, 07.05.Tp, 66.20.+d}

\section{Introduction}
The simulation of the dynamics of colloidal suspensions is a challenging
task. The reason is that the movement of the colloidal particles can be
on a time scale which is orders of magnitudes slower than that of the
solvent particles (e.g.~seconds versus picoseconds). Therefore,
simulation methods such as Molecular Dynamics, that account for the
fluid particles explicitly, are not well suited to study the dynamics of
colloidal suspensions because one would spend most of the simulation time
solving the equations of motion of the fluid particles. The situation for
charged colloids is even worse because one has to take into account the
long--ranged Coulomb interactions which are already very time consuming
in simple ionic liquids.

One possibility to circumvent these problems is to avoid explicit
simulation of the solvent particles and describing the interaction
between the colloidal particles by means of an effective potential. In
the case of charged colloidal systems this is usually a Yukawa--like
potential which gives in many cases quite an accurate description
of interaction--dependent properties~\cite{stevens96,alla98}. But
the approach with effective interactions neglects completely the
hydrodynamic interactions between the colloidal particles which stem
from the momentum transport through the solvent. In order to take
this into account one has to treat the hydrodynamics of the solvent
at least on a coarse--grained level. An effective scheme, that was
developed to solve efficiently the Navier--Stokes equations, is the
so--called lattice Boltzmann method (LBM). The LBM is a pre--averaged
version of a lattice gas, i.e., a Boltzmann equation is solved on a
lattice such that the Navier--Stokes equations are recovered (reviews
of the method are the Refs.~\cite{benzi92,ladd94,chen98}).  Recently,
the LBM was applied to simulate the dynamics of colloidal systems, such
as the rotational and translational short--time dynamics of colloids
\cite{ladd93,hagen98,hagen99,lowe95,heemels00}, the diffusion of colloidal
particles in confined geometry~\cite{hagen97,pago98,pago99}, and the
dynamics in porous media~\cite{oxaal94,koponen98,lowe96,koch98}. Also
other complex systems like polymer solutions \cite{ahlrichs99} have been
investigated by LBM.

Several attempts to apply the LBM to charged systems have been
reported in the literature. He and Li~\cite{he00} proposed a scheme
which is appropriate to  study electrochemical processes in an
electrolyte. However, in this method it is assumed that the fluid
is {\it locally} electrically neutral which cannot be true for the
part of an electrolyte forming the electrical double layer around
a macroion. Thus, the LBM of He and Li cannot be used to describe
the dynamics of suspensions of charged colloids. A different LBM for
charged systems was suggested by Warren~\cite{warren97}. The central
idea of this method is the introduction of external charge densities
$\rho_s$ for the ionic species of type $s$ which are propagated with
the one particle distribution function of the LB equation by means of
the so--called moment propagation method~\cite{frenkel89}. These ionic
densities are coupled back to the mass current of the LB equation via
a chemical potential which consists of a term $\propto {\rm ln} \rho_s$
and a term proportional to the electrostatic potential such that it only
gives a contribution if the ion densities are not distributed as expected
in equilibrium by a Boltzmann distribution (e.g.~because of an external
electrical field). Finally, the electrostatic potential is determined
from the charge densities by means of a Poisson equation solver. The main
drawback of Warren's method is that the charge densities are introduced
as additional physical quantities which are independent from the mass
density in the LB equation, and thus, this method is not self--consistent
and the correctness of coupling of the charge density to the mass current 
is not obvious.

Our LBM for charged colloidal suspensions is inspired by Warren's
approach, but it does not introduce ionic species as additional
quantities. The details of our method can be found in the following
sections. We apply it to the determination of the sedimentation velocity
of an array of charged spheres in an electrolyte.

The article is organized as follows: In Sec.~II we give a brief
introduction into the electrokinetic equations of motion with which we
model the dynamics of the fluid. In Sec.~III the LBM is presented which
solves the latter equations of motion on a lattice. The LBM is verified
in Sec.~IV by means of a Chapman--Enskog expansion. And in Sec.~V we
show our results for the sedimentation phenomena. We conclude with a
discussion of the method and the results.

\section{The electrokinetic equations}
In this section we introduce the equations of motion for a hydrodynamic
description of charged colloidal suspensions. These equations can be
found in standard textbooks (see e.g.~Ref.~\cite{hunter}).

Consider a system of macroions with radius $a$ in an electrolyte
consisting of two ionic species which have charges $+z_1 e$ and $-z_2
e$, respectively (where $z_1$ and $z_2$ are the valencies of the
ions and $e$ is the charge of a proton). The densities of the ions,
$\rho_1(\vec{r},t)$ and $\rho_2(\vec{r},t)$, are conserved quantities
and therefore each of them follows a continuity equation:
\begin{equation}
  \frac{\partial \rho_s} {\partial t} =
  - \nabla \cdot \vec{J}_s \ \ \ \ \ \ \ \ \ \ s=1,2 \ .
  \label{eq1}
\end{equation}
The current $\vec{J}_s$ is given by
\begin{equation}
   \vec{J}_s = 
     \rho_s \vec{u}
     - D_s \nabla \rho_s
     - z_s D_s \rho_s \nabla \hat{\Phi} \ .
    \label{eq2}
\end{equation} 
The first term in (\ref{eq2}) in which $\vec{u}$ denotes the flow velocity
is the convection current whereas the two other terms describe the
diffusive current and the current due to the electrostatic potential $\Phi$.
$D_s$ denotes the diffusivity of ions of type $s$ and
$\hat{\Phi}$ is the electrostatic potential in dimensionless form,
$\hat{\Phi} = \frac{e}{k_B T} \Phi$ ($k_B$: Boltzmann constant, 
$T$: temperature). 

$\hat{\Phi}$ is determined by the Poisson equation,
\begin{equation}
  \nabla^2 \hat{\Phi} = - 4 \pi l_B \left( \sum_{s=1}^2 z_s \rho_s
					   + \sigma \right)
  \label{eq3}
\end{equation}
where the Bjerrum length $l_B$ is defined by
\begin{equation}
  l_B = \frac{e^2}{4 \pi \epsilon k_B T} \ .
  \label{eq4}
\end{equation}
$\epsilon$ is the dielectric constant. At a distance $l_B$ the Coulomb energy
of one ion due to another ion is equal to $k_B T$. $\sigma$ denotes the charge density
of the macroion. We assume that the charge $Z$ of each macroion sits on its surface
in the form of uniformly distributed point charges.

If we set $\vec{u}=0$ in Eqs.~(\ref{eq1}) and (\ref{eq2}) we yield the
equilibrium solution for the ion densities as
\begin{equation}
  \rho_s (\vec{r}) = 
  \overline{\rho}_s \exp \left(- z_s \hat{\Phi} (\vec{r}) \right) \ . 
  \label{eq5}
\end{equation}
By putting the Boltzmann distribution (\ref{eq5}) into the Poisson
equation (\ref{eq3}) this leads to the so--called Poisson--Boltzmann equation
in which correlations between the ions are neglected. Moreover, if one 
linearizes the exponential function in Eq.~(\ref{eq5}) (Debye--H\"uckel
theory) it is possible to solve Eq.~(\ref{eq3}) analytically and the result is
a Yukawa potential,
\begin{equation}
  \hat{\Phi} (\vec{r}) = K 
  \frac{\exp \left( - |\vec{r}|/\lambda_D \right)}{|\vec{r}|} \ .
   \label{eq6}
\end{equation}
The so--called Debye length is defined by
\begin{equation}
   \lambda_D \equiv \kappa^{-1} = \frac{1}{\sqrt{4 \pi l_B \sum_s z_s^2
                        \overline{\rho}_s}} \ .
   \label{eq7}
\end{equation}
The potential (\ref{eq6}) is even a good approximation in the non--linear
case but the prefactor $K$ is then different from the one of the linear
theory. 

The equation of motion which has to be specified finally is that for the total mass 
current of the fluid, $\rho \vec{u} \equiv (\sum_{s=1}^{2} \rho_s + \rho_n) \vec{u}$,
where $\rho_n$ is the density of the neutral part of the fluid.
We assume that our fluid can be described by the linearized Navier--Stokes equations for 
low Reynolds number flow. Hence, the equation for $\rho \vec{u}$ with a body force 
due to the electrostatic potential is
\begin{equation}
    \frac{ \partial (\rho \vec{u})}{\partial t} =
    \nu \nabla^2 \rho \vec{u} - \nabla p 
    - k_B T \sum_s z_s \rho_s \nabla \hat{\Phi}
    \label{eq8}
\end{equation}
where $p$ is the pressure and $\nu$ is the kinematic viscosity. If the equation
for the pressure is the one of an ideal gas, $p=k_B T \rho$, $p$ can be decomposed 
into an electrostatic and a neutral part, $p_e= k_B T \sum_{s=1}^2 \rho_s$ and
$p_n=k_B T \rho_n$, respectively. The sum of $- \nabla p_e$ and the electrostatic 
body force, i.e.~the last term in Eq.~(\ref{eq8}), is zero, if the ion densities have 
relaxed to their equilibrium distribution, Eq.~(\ref{eq5}).

\section{The Lattice--Boltzmann method for charged colloids}
We have developed a simulation method to solve the non--linear coupled
Eqs.~(\ref{eq1}), (\ref{eq3}), and (\ref{eq8}) numerically. For this we 
use concepts which are well--known from the LBM.

In this method the discretized version of a Boltzmann equation is solved 
numerically on a lattice on which every lattice point represents a cell of
particles. The central quantity is the one--particle distribution function 
$n_i(\vec{r},t)$ which describes the number of particles on a lattice node 
$\vec{r}$ at time $t$ with a discrete velocity $\vec{c}_i$. The discrete 
space of velocities $\{ \vec{c}_i \}$ is chosen such that no artificial
anisotropic terms appear in the corresponding equations in the continuous limit. 
In our case the velocity space consists of 18 vectors of which from a given lattice
node 6 point to the nearest and 12 to next--nearest neighbors on a simple 
cubic lattice. This velocity space can be constructed by projecting the
unit vectors of a four dimensional FCHC lattice onto three dimensions. 
It is one possible choice of a velocity space which exhibits
the required isotropy. 

The equation of motion for $n_i(\vec{r},t)$ consists of two steps, a collision 
and a propagation step. In the collision step the interaction between the particles 
is taken into account which results in the post--collision function 
$n_i^{\star}(\vec{r},t^{\star})$ at the collision time $t^{\star}$. In
the propagation step $n_i(\vec{r},t)$ is updated by
\begin{equation}
   n_i(\vec{r}+\vec{c}_i,t+1) = n_i^{\star}(\vec{r},t^{\star}) \ .
   \label{eq9}
\end{equation}
In this equation the lattice constant, the time step, and the mass of a
particle is set to unity. The density $\rho(\vec{r},t)$ and the mass
current $\vec{j} \equiv \rho \vec{u}$ are given by the zeroth and first
moment of $n_i$, respectively, $\rho = \sum_i n_i(\vec{r},t)$, $\vec{j}
= \sum_i n_i(\vec{r},t) \vec{c}_i$.

In the case of the charged system a one--particle distribution function
for each ion species and a neutral part is required. The purpose of the
neutral part is to keep the viscosity essentially constant through the
fluid.  Thus, it is chosen such that its value at a given lattice point
is much higher than that of the ionic densities. We make the following
Ansatz for the post--collision function $n_i^{s \star}$ for the counter--
and coions, $s=+,-$, respectively, and the neutral part, $s=n$ (we also
take into account rest particles by the index $i=0$):
\begin{eqnarray}
  n_i^{s \star} (\vec{r},t^{\star}) & = &
  \frac{{\rm w}_i \gamma_s }{24} \rho_s (\vec{r},t) 
  \left(
     1 + \frac{1}
              {c_{{\rm sv}}^2 \rho^{\prime}(\vec{r},t)}
              \vec{j}(\vec{r},t) \cdot \vec{c}_i  \right) \ ,
             \label{eq10} \\
  n_0^{s \star} (\vec{r},t^{\star}) & = &
              \left(1-\gamma_s \right) \rho_s (\vec{r},t) \ .
              \label{eq11}
\end{eqnarray}
The factor ${\rm w}_i$ is a weighting factor which is equal to 2 for the
$\vec{c}_i$ in the direction of nearest neighbors and equal to 1 for the
remaining $\vec{c}_i$. So it satisfies the normalization constraint 
$\sum_{i=1}^{18} \frac{{\rm w}_i}{24} = 1$. For the following we define also ${\rm w}_0:=0$.
With the parameter $0 < \gamma_s \le 1$ the diffusivity $D_s$ of the particles of type 
$s$ can be varied. The latter quantity is given by
\begin{equation}
  D_s = \frac{c_{{\rm sv}}^2}{2} \gamma_s
  \label{eq12}
\end{equation}
which is shown in the next section. $c_{{\rm sv}}$ is the sound velocity 
which is $1/\sqrt{2}$ for our model~\cite{ladd94}. The density $\rho^{\prime}$ 
is defined by $\rho^{\prime}= \sum_s \gamma_s \rho_s$.

The propagation step for our charged system is
\begin{eqnarray}
  n_i^s (\vec{r}+\vec{c}_i,t+1)& = & n_i^{s \star} (\vec{r},t^{\star}) ,
  \label{eq13} \\
  n_0^s(\vec{r},t+1) & = & n_0^{s \star} (\vec{r},t^{\star}) \ .
  \label{eq14}
\end{eqnarray}
Different propagation rules have to be established at the surface of the
macroions and at walls. Here we use the bounce back rules suggested
by Ladd \cite{ladd94} which lead to no--slip boundary conditions. In
this scheme one puts a sphere which represents a macroion onto the
lattice whereby its surface cuts links between lattice nodes. The
boundary nodes are defined halfway along these links and the population
functions $n_i^s$ which point to the direction of the boundary nodes
are reflected back during the propagation step. In the case of moving
boundaries there is a momentum transfer between the boundary nodes and
the fluid. In this case the self consistent scheme derived by Lowe {\it
et al.}~\cite{lowe95,heemels00} can be used.

The aforementioned way of mapping a sphere onto the lattice introduces
fluid inside and outside the sphere. In our scheme we assign the charge of
the macroions in that the inner fluid is an electrolyte with net charge
$Z$. Charge neutrality requires that the total charge of the outer
fluid equals the one of the inner fluid of the macroions. Of course,
it is not allowed in our scheme that outer fluid leaks through the
surface of the sphere.  Therefore, only small movements of a macroion
are possible such that the center of mass of the sphere can be fixed,
and only a momentum transfer with the fluid takes place.

The densities $\rho_s$ and the total mass current $\vec{j}$ cannot be inferred 
simply from the zeroth and first moments of the $n_i^s$'s because we have to 
take into account their coupling to the gradient of the electrostatic potential. 
If $\rho_s$ and $\vec{j}$ are calculated as follows,
\begin{equation}
  \rho_s (\vec{r},t+1) = \sum_{i=0}^{18} \left(
       n_i^s (\vec{r},t+1) 
       - \frac{z_s}{2} \frac{{\rm w}_i \gamma_s}{24}
         \rho_s(\vec{r}-\vec{c}_i,t) 
          \nabla \hat{\Phi} (\vec{r}-\vec{c}_i,t) \cdot \vec{c}_i
       \right)
  \label{eq15}
\end{equation}
and 
\begin{equation}
   \vec{j}(\vec{r},t+1) = \sum_{s=1}^{3} \left( \sum_{i=1}^{18} 
     n_i^s (\vec{r},t+1) \vec{c}_i
     - c_{{\rm sv}}^2 z_s \gamma_s \rho_s(\vec{r},t+1)
       \nabla \hat{\Phi} (\vec{r},t+1) \right) \  .
    \label{eq16}
\end{equation}
we are consistent with Eqs.~(\ref{eq1}) and (\ref{eq8}) in the continuous limit.
$\nabla \hat{\Phi}$ does not couple to the neutral part of the fluid.
This is guaranteed in Eqs.~(\ref{eq15}) and (\ref{eq16}) by setting $z_s=0$
for $s=n$.

If one replaces $n_i^s (\vec{r},t+1)$ by $n_i^s (\vec{r}-\vec{c}_i,t)$
in Eqs.~(\ref{eq15}) and (\ref{eq16}) by using Eq.~(\ref{eq13}) it
becomes clear that Eqs.~(\ref{eq15}) and (\ref{eq16}) are the discrete
versions of Eqs.~(\ref{eq1}) and (\ref{eq8}), respectively. Thereby, the
second terms in Eqs.~(\ref{eq15}) and (\ref{eq16}) correspond respectively
to the current due to the electrostatic potential in Eq.~(\ref{eq1})
and the electrostatic body force in Eq.~(\ref{eq8}). We show this
explicitly in the next section by means of a Chapman--Enskog expansion.
Of course, with the $\nabla \hat{\Phi}$ terms in Eqs.~(\ref{eq15})
and (\ref{eq16}) the conservation of the ionic densities and the mass
current is still fulfilled.  This is guaranteed because $\hat{\Phi}$
is determined self--consistently from the ionic densities by means of
the Poisson equation.

This means that we have to solve the Poisson equation at each time step in order
to determine the electrostatic potential from the ionic densities $\rho_+$
and $\rho_-$.  For this purpose we use a successive over--relaxation
(SOR) scheme in which one looks in principle for the stationary solution
of a diffusion equation \cite{num_rec}. But in contrast to a normal
diffusion equation one introduces an acceleration parameter $1<\omega<2$
such that the potential is obtained as the iterative solution of the
following equation:
\begin{equation}
  \hat{\Phi}_{h+1} (\vec{r},t) =
  \omega \left[ \sum_i \left( \frac{{\rm w}_i}{24} 
                              \hat{\Phi}_h(\vec{r}-\vec{c}_i,t)
                             + 4 \pi l_B \frac{c_{{\rm sv}}^2}{2} 
                               \sum_{s=+,-} z_s \rho_s(\vec{r},t)
                       \right) \right]
     + (1-\omega) \hat{\Phi}_h (\vec{r},t)  \ .
 \label{eq17}
\end{equation}
In this equation we have denoted the iteration time by $h$ (the unit for an 
iteration step is again set to unity). For a sufficient number of iteration
steps $\Phi_h$ converges to $\Phi$. We have found that a suitable choice 
for $\omega$ is $1.45$ which guarantees stability and optimal acceleration. 
The gradient of $\hat{\Phi}$ which we need in Eqs.~(\ref{eq15}) and 
(\ref{eq16}) is given by 
\begin{equation}
   \nabla \hat{\Phi} (\vec{r}) = - \sum_i \frac{{\rm w}_i}{24 c_{{\rm sv}}^2}
    \hat{\Phi} (\vec{r}-\vec{c}_i) \vec{c}_i \ .
    \label{eq18}
\end{equation}
Note that we use in Eqs.~(\ref{eq17}) and (\ref{eq18}) the same discrete space 
as the velocity space in the Lattice--Boltzmann equations. This means that the 
truncation error which is caused due to the discrete representation of the 
derivatives is of fourth order in contrast to a second order truncation error 
on a simple cubic lattice with 6 vectors $\{\vec{c}_i\}$ pointing to the 
nearest neighbors from a given lattice node. The stability and efficiency
of the SOR algorithm are further optimized by making use of a partially 
decoupled red--black Gauss--Seidel scheme \cite{gupta97}. 

\section{Chapman--Enskog expansion}
By means of a Chapman--Enskog expansion we show in this section that the
set of discrete equations from the previous section indeed recover the
Eqs.~(\ref{eq1}), (\ref{eq3}), and (\ref{eq8}) in the continuous limit.

As a first step we rewrite Eqs.~(\ref{eq15}) and (\ref{eq16}) for the 
densities $\rho_s$ and the partial currents 
$\vec{j}_s \equiv \gamma_s \rho_s \vec{j} / \rho^{\prime}$:
\begin{eqnarray}
  \rho_s(\vec{r},t) & = & \sum_i \frac{{\rm w}_i \gamma_s}{24}
                    \left( 
                      \rho_s(\vec{r}-\vec{c}_i,t-1) 
                     + \vec{j}_s(\vec{r}-\vec{c}_i,t-1) \cdot \vec{c}_i
                    \right. \nonumber \\
                  & & \left. + \frac{z_s}{2}
                      \rho_s(\vec{r}-\vec{c}_i,t-1) 
                     \nabla \hat{\Phi} (\vec{r}-\vec{c}_i,t-1) \cdot \vec{c}_i 
                      \right) \ , \label{eq19} \\
  \vec{j}_s(\vec{r},t) & = & \sum_i \frac{{\rm w}_i \gamma_s}{24}
                    \left( 
                      \rho_s(\vec{r}-\vec{c}_i,t-1) \vec{c}_i 
                     + \vec{j}_s(\vec{r}-\vec{c}_i,t-1) \cdot \vec{c}_i \vec{c}_i
                    \right) \nonumber \\
                    & & - c_{{\rm sv}}^2 \gamma_s z_s
                      \rho_s(\vec{r},t) \nabla \hat{\Phi}(\vec{r},t) \ .
                      \label{eq20}
\end{eqnarray}
If we now expand the functions of the form $f(\vec{r}-\vec{c}_i,t-1)$ 
($f=\rho_s,\vec{j}_s,\rho_s \nabla \hat{\Phi}$) around position $\vec{r}$ and 
time $t$ up to second order, 
\begin{equation}
f(\vec{r}-\vec{c}_i,t-1) = f(\vec{r},t) + 
                          \left( -\partial_t 
                          - c_{i \alpha} \nabla_{\alpha} + \frac{1}{2} 
                          \left(
                           \partial_t + c_{i \alpha} \nabla_{\alpha}
                          \right) ^2 
                          \right) f(\vec{r},t) \ , 
                    \label{eq21}
\end{equation}
we obtain the following equations for $\rho_s$ and $\vec{j}_s$,
\begin{eqnarray}
  \partial_t \rho_s & = & \frac{1}{2} \partial_t^2 \rho_s 
                         - \nabla \cdot \vec{j}_s
                         + \partial_t \nabla \cdot \vec{j}_s
        + \frac{c_{{\rm sv}}^2}{2} \gamma_s \left(
                          \nabla^2 \rho_s
                         + z_s \nabla \cdot \rho_s \nabla \hat{\Phi} 
                         + z_s \partial_t \nabla \cdot \rho_s \nabla \hat{\Phi}
                           \right)
                          \label{eq22}  \\
  \partial_t \vec{j}_s & = & - c_{{\rm sv}}^2 \gamma_s \nabla \rho_s 
                             + c_{{\rm sv}}^2 \gamma_s \partial_t \nabla \rho_s
                             + \frac{1}{2} \partial_t^2 \vec{j}_s
                             + \frac{1}{6} \nabla^2 \vec{j}_s
                             - c_{{\rm sv}}^2 \gamma_s z_s \rho_s 
                               \nabla \hat{\Phi} \ .
                        \label{eq23}
\end{eqnarray}
To achieve Eqs.~(\ref{eq22}) and (\ref{eq23}) we have used the 
lattice sums~\cite{ladd94} 
\begin{eqnarray}
   \sum_{i=1}^{18} c_{i \alpha} c_{i \beta} & = &
                c_{{\rm sv}}^2 \delta_{\alpha \beta} \  , \label{eq24} \\ 
   \sum_{i=1}^{18} c_{i \alpha} c_{i \beta} c_{i \gamma} c_{i \delta} & = &
             \frac{1}{3} \left(
               \delta_{\alpha \beta} \delta_{\gamma \delta} +
               \delta_{\alpha \gamma} \delta_{\beta \delta} +
               \delta_{\alpha \delta} \delta_{\beta \gamma} 
              \right) \ .
         \label{eq25}
\end{eqnarray}

Eqs.~(\ref{eq22}) and (\ref{eq23}) are the starting point for the 
Chapman--Enskog expansion which introduces a macroscopic space
scale by $\vec{r}_1 = \epsilon \vec{r}$ and two macroscopic time
scales by $t_1= \epsilon t$ and $t_2 = \epsilon^2 t$. The $t_1$
scale describes fast convection processes whereas on the slower $t_2$
scale the diffusion of vorticity takes place. Thus, the Chapman--Enskog
expansion enables us to consider the latter time scales separately.
It was for the first time applied to lattice gases by 
Frisch {\it et al.}~\cite{frisch86}.

We now express the derivatives by means of $\vec{r}_1$, $t_1$, and $t_2$,
\begin{eqnarray}
  \nabla & = & \epsilon \nabla_1 \label{eq26} \\
  \partial_t & = & 
    \epsilon \partial_{t_1} + \epsilon^2 \partial_{t_2} \ , \label{eq27}
\end{eqnarray}
and put them into Eqs.~(\ref{eq22}) and (\ref{eq23}). If we collect terms of the
same order in $\epsilon$ we obtain on the $\epsilon^1$ scale
\begin{eqnarray}
   \partial_{t_1} \rho_s & = & - \nabla_1 \cdot \vec{j}_s \ , \label{eq28} \\ 
   \partial_{t_1} \vec{j}_s & = & - c_{{\rm sv}}^2 \gamma_s
                 \left( \nabla_1 \rho_s +z_s \rho_s \nabla_1 \hat{\Phi} 
                 \right)  \ . 
          \label{eq29}
\end{eqnarray}
Eq.~(\ref{eq28}) is the continuity equation for mass conservation. If one
takes the sum over $s$ on both sides of Eq.~(\ref{eq29}) one obtains
the "fast" part of the linearized Navier--Stokes equation for the total mass current
$\vec{j}$:
\begin{equation}
  \partial_{t_1} \vec{j} = - c_{{\rm sv}}^2 \sum_s \left(\nabla \rho_s 
			 + z_s \rho_s \nabla_1 \hat{\Phi} \right) \ .
   \label{eq29a}
\end{equation}
The first term on the right--hand side of this equation is the negative gradient of
the pressure and the second term the electrostatic body force. 

On the $\epsilon^2$ scale we have
\begin{eqnarray}
  \partial_{t_2} \rho_s & = & \frac{1}{2} \partial_{t_1}^2 \rho_s
                       + \partial_{t_1} \nabla_1 \cdot \vec{j}_s
                   + \frac{c_{{\rm sv}}^2}{2} \gamma_s \left(
                      \nabla_1^2 \rho_s + z_s
                      \nabla_1 \cdot \rho_s \nabla \hat{\Phi}
                      \right) \ , \label{eq30} \\
  \partial_{t_2} \vec{j}_s & = & c_{{\rm sv}}^2 
		     \gamma_s \partial_{t_1} \nabla_1 \rho_s
                   + \frac{1}{2} \partial_{t_1}^2 \vec{j}_s 
                   + \frac{1}{6} \nabla^2 \vec{j}_s \ . 
            \label{eq31}
\end{eqnarray}

From Eqs.~(\ref{eq28})--(\ref{eq31}) we see that the transport of the ion densities
to equilibrium can be either achieved by convection or by diffusion. So, if
$\partial_{t_1} \vec{j}_s$ vanishes Eq.~(\ref{eq29}) is solved by the Boltzmann
distribution, Eq.~(\ref{eq5}). And   
by using Eqs.~(\ref{eq28}) and (\ref{eq29}) Eq.~(\ref{eq30}) for the densities 
simplifies to
\begin{equation}
  \partial_{t_2} \rho_s = 0 \ . \label{eq32}
\end{equation}
This equation implies that the fluid is incompressible on the $t_2$ scale. 
In this case there is no diffusion on the $t_2$ scale because the ionic densities
have already come to their equilibrium on the faster $t_1$ scale. On the other
hand, if we assume that the second derivative of $\rho_s$ with respect to
$t_1$ vanishes then Eq.~(\ref{eq30}) becomes a diffusion equation
with the diffusion constant $D_s = c_{{\rm sv}}^2 \gamma_s/2$ for ions 
of type $s$. The diffusion process relaxes the densities $\rho_s$ ($s=+,-$)
again to the equilibrium distribution (\ref{eq5}).

We still have to discuss Eq.~(\ref{eq31}) for the mass current on the
$t_2$ scale. It is reasonable to assume that the derivatives of $\rho_s$
and $\vec{j}_s$ with respect to $t_1$ are small on the $t_2$ scale.
So we may neglect the first two terms in Eq.~(\ref{eq31}). Furthermore, we
have to sum over $s$ on both sides of Eq.~(\ref{eq31}) in order to obtain
the equation of motion for the total current $\vec{j}$ on the $t_2$ scale:
\begin{equation}
   \partial_{t_2} \vec{j} = \frac{1}{6} \nabla_1^2 \vec{j} \ . 
     \label{eq33}
\end{equation}
Moreover, if we combine this equation with Eq.~(\ref{eq29}) for variations on
the $t_1$ scale we obtain the following equation:
\begin{equation}
  \partial_{t} \vec{j} = \frac{1}{6} \nabla^2 \vec{j} 
                         - c_{{\rm sv}}^2 \sum_s \gamma_s \nabla \rho_s
                         - c_{{\rm sv}}^2 \sum_s \gamma_s z_s
                           \rho_s \nabla \hat{\Phi} \ .
        \label{eq34}
\end{equation}
So we recover Eq.~(\ref{eq8}) for the total current whereby the kinematic 
viscosity $\nu$ is $1/6$ and the equation for the pressure depends on the 
parameters $\gamma_s$,
\begin{equation}
  p = c_{{\rm sv}}^2 \sum_s \gamma_s \rho_s \ .
  \label{eq35}
\end{equation}
Note that for $\gamma_1=\gamma_2$ this is just the equation of state for an ideal
gas.

\section{The sedimentation velocity}
In this section we present the results for the sedimentation velocity of
an array of charged spheres in an electrolyte solution. We show that our
method gives correct results in that it recovers an analytical formula
which is valid in the limit of an isolated, weakly charged sphere in an
unbounded electrolyte.  Furthermore, we discuss the dependence of the
sedimentation velocity on the charge of the spheres.  We then compare
our results to experimental data.

Before we show the results for the sedimentation velocity we discuss
to what extent the calculated electrostatic potentials and the ionic
densities around a macroion are influenced by lattice artifacts. The
latter may have several reasons: We use a simple way to introduce the
charge of a macroion. We assign its charge $Z$ by distributing the
densities on the lattice nodes inside the macroion such that the net
charge, i.e.~the sum over all these nodes, is equal to $Z$. This may
have the effect that the effective charge distribution on the surface
of the macroion is anisotropic because the inner nodes form only an
approximative representation of a sphere which can, of course, be
improved by increasing its radius $a$. Another drawback which is due
to the lattice lies in the range of the electrostatic potential. The
Debye length, measured in lattice units, must be at least larger than
one. Otherwise, the simulation becomes unstable because one has strong
discontinuities in the ionic densities between two lattice points close
to the surface of the macroion.

\begin{figure}[h]
\psfig{file=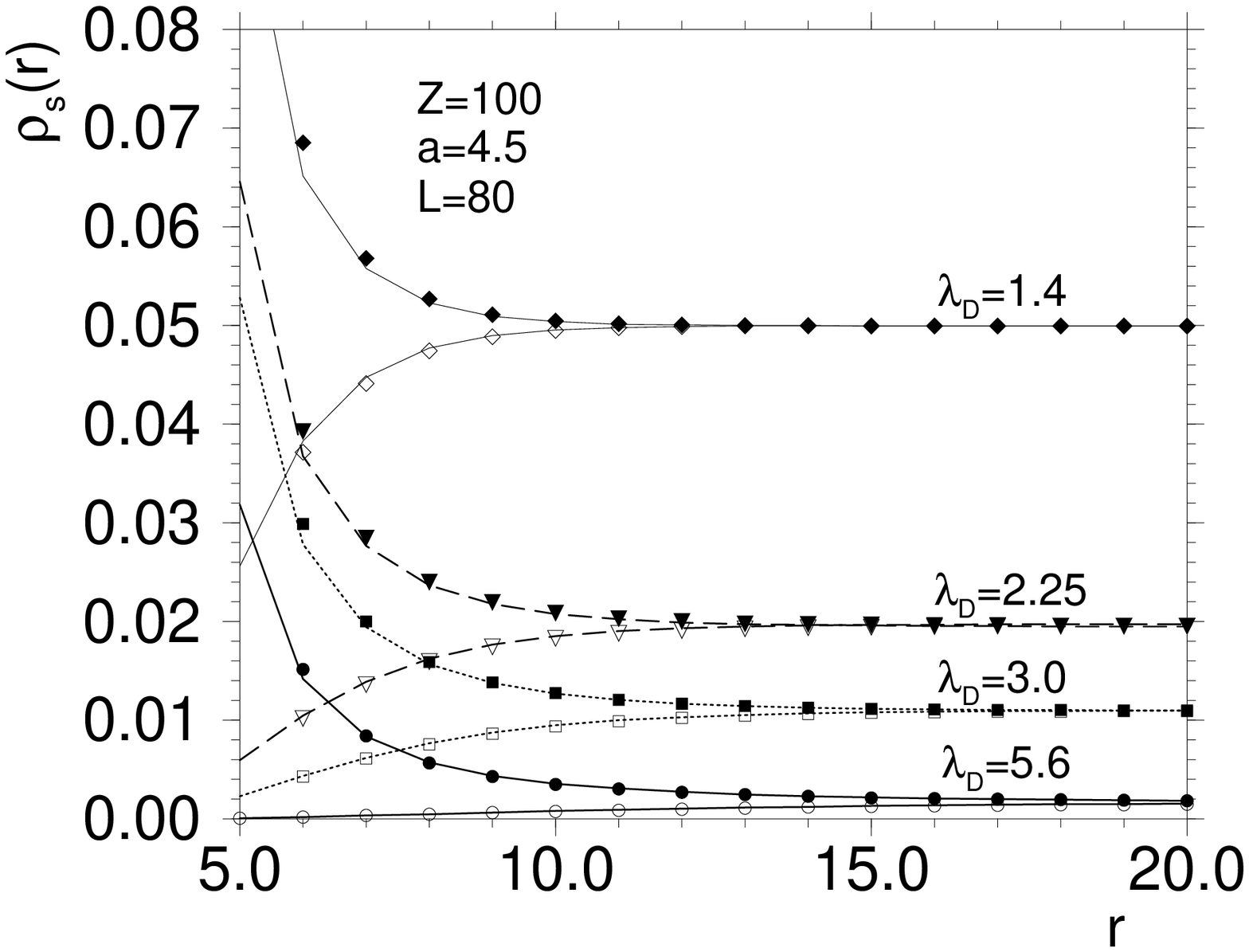,width=16cm,height=13cm}
\vspace{4mm}

\caption[fig1]{The radial ionic densities around a macroion with
               $a=4.5$ for $Z=100$ as determined directly from 
               $\rho_s(\vec{r})$ (filled symbols for the counterions
               and open symbols for the coions) and by calculating it
	       from the electrostatic potential (lines) --- see text ---
	       for the indicated Debye lengths $\lambda_D$.}
\label{fig1}
\end{figure}

The importance of these artifacts can be checked by determining the
radial ionic densities $\rho_s(r)$ around a spherical macroion, i.e.~the
densities for the ions of type $s$ at a distance $r$ from the center of
the sphere.  In the absence of artifacts $\rho_s(r)$ should obey
\begin{equation}
 \rho_s(r) = \overline{\rho}_s \exp \left( - z_s \hat{\Phi}(r) \right) \ .
 \label{eq_phi}
\end{equation}
In Fig.~\ref{fig1} we show an example of the $\rho_s(r)$ calculated
directly (symbols) and from the right--hand side of Eq.~(\ref{eq_phi})
(solid lines) for a macroion with a radius $a=4.5$ and charge $Z=100$. The
length of the simulation box is $L=80$ and the density of the neutral
fluid is set to $\rho_n=20$.  The Debye length is varied by changing the
sum of the mean ionic densities from $\sum_s \overline{\rho}_s=0.0063$
($\lambda_D=5.6$) to $\sum_s \overline{\rho}_s=0.1$ ($\lambda_D=1.4$)
whereby the Bjerrum length is $l_B=0.4$.  We can infer from
Fig.~\ref{fig1} that the ionic densities as calculated directly agree very
well with those calculated from the right--hand side of Eq.~(\ref{eq_phi})
even for a Debye length as small as $\lambda_D=1.4$. This means that
$a=4.5$ is a reasonable choice for the radius of a macroion which keeps
lattice artifacts small at least for $\lambda_D \ge 1.4$.

Now we consider one charged macrosphere in a unit cell of length $L$
with a fixed position. It represents one particle in an array of spheres
on a simple cubic lattice because it interacts with its own periodic
images. The volume fraction of the macroions is given by $\varphi=(4 \pi
a^3)/(3 L^3)$. In order to investigate sedimentation phenomena we have
to introduce a gravitational force in the LB equations. After a time
of the order $\tau_{{\rm s}}=L^2/\nu_{{\rm eff}}$ ($\nu_{{\rm eff}}$:
effective, kinematic viscosity of the fluid in the presence of the
spheres) the steady state is reached for which one determines the average
flow velocity in the unit cell. We divide the latter by the average flow
velocity of the corresponding neutral system and yield the ratio $U/U_0$
of the sedimentation velocities in the charged and the neutral system,
respectively. For the larger systems considered, we did not wait until the
system came to its steady state, because we know that the time dependence
of the apparent sedimentation velocity $U_{{\rm app}}(t)$ is given by
\begin{equation}
  U_{{\rm app}}(t) = U 
  \left( 1 - \exp \left( - \frac{t}{\tau_{{\rm s}}} \right) \right) \ .
  \label{uapp}
\end{equation}
By computing numerically the time derivative of Eq.~(\ref{uapp}) one
obtains a simple exponential function with the two unknown quantities
$U$ and $\tau_s$ which we computed from fits of the logarithm of this
exponential function. With this procedure it was possible to determine
$U$ within a time of the order $t \approx \tau_{{\rm s}}/20$.  We checked
the accuracy of our fits at $\varphi=0.0018$ by comparing them to the
exact steady state results, and we obtained identical results for $U/U_0$
as a function of $\kappa a$ ($\kappa \equiv \lambda_D^{-1}$).

In the following we show that our LB method recovers an
analytical result for $U/U_0$ which was first derived by
Booth~\cite{booth54}, and later slightly modified by Ohshima {\it et
al.}~\cite{ohshima84}.\footnote{However, Booth's result agrees with
the one of Ohshima {\it et al.}~for an 1--1--electrolyte.  We are only
interested in this special case in this paper.} It is valid in the limit
of infinite dilution and small charge of the macroions, i.e.~a weakly
charged macroion in an electrolyte with infinite extension. What do we
expect in this case? Due to the external force the ionic concentrations
around the macroions which form the electrical double layer deviate
from their equilibrium values. The double layer loses its spherical
symmetry due to the fluid motion which results in an electrical dipole
field pointing in the direction opposite to the motion of the macroion
and thus reduces its sedimentation velocity.  Booth's calculation starts
with the Ansatz
\begin{equation}
  \frac{U}{U_0} = 1 + \sum_{k=1}^{\infty} c_k Z^k
\end{equation}
and takes into account only terms in the lowest non--vanishing order in $Z^k$,
\begin{equation}
  \frac{U}{U_0} = 1 + c_2 Z^2 \  .
  \label{appbooth}
\end{equation}
The coefficient $c_2$ can be calculated analytically by solving
the electrokinetic equations of motion (\ref{eq1}), (\ref{eq3}),
and (\ref{eq8}) whereby the Poisson equation is solved in the
Debye--H\"uckel limit. The final expression for $c_2$ has the following
form~\cite{booth54,ohshima84}:
\begin{equation}
   c_2 = - \frac{k_B T l_B}{72 \pi a^2 \eta} 
        \frac{\sum_s z_s^4 \overline{\rho}_s D_s^{-1}}{\sum_s z_s^2 
        \overline{\rho}_s} \; f(\kappa a) \ .
\end{equation}
$\overline{\rho}_s$ denotes the mean density of the ions of type $s$ far away 
from the center of the macroion. $\eta$ is the shear viscosity. $f(\kappa a)$ is 
a function of exponential integrals of different order $n$, 
${\rm E}_n (x) = x^{n-1} \int_x^{\infty} dt \; t^{-n} \exp(-t)$:
\begin{eqnarray}
   f(\kappa a) & = & \frac{1}{1+(\kappa a)^2} 
                 \left[ {\rm e}^{2 \kappa a} 
                        \left( 3 E_4 (\kappa a) - 5 E_6 (\kappa a) \right)^2
                       + 8 {\rm e}^{\kappa a} 
                         \left(E_3(\kappa a) - E_5(\kappa a) \right) \right. 
                     \nonumber \\
               & &   \left.     - {\rm e}^{2 \kappa a}
                         \left( 4 E_3(2 \kappa a) + 3 E_4(2 \kappa a) 
                               - 7 E_8(2 \kappa a) \right)
                 \right]  \ . 
\end{eqnarray}
We have determined $U/U_0$ as a function of $\kappa a$ for
different volume fractions. The radius of the macroion is fixed to
$a=4.5$. Moreover, the diffusivities of the ionic species in the fluid
are chosen to be $D_1=0.165$ for the counterions and $D_2=0.25$ for
the coions. The ratio $D_1/D_2=0.66$ corresponds to that of $D_{{\rm
Na}}/D_{{\rm Cl}}$ in sodium chloride. The Bjerrum length is again set
to $l_B=0.4$.  In order to vary $\kappa a$ from $0.15$ to $1.5$ we
have to change $\sum_s \overline{\rho}_s$ from $0.00025$ to $0.02211$,
respectively. This is small compared to the density of the
neutral fluid, $\rho_n=20$. So by changing $\kappa a$ we do not change
the viscosity of the fluid significantly.

\begin{figure}[h]
\psfig{file=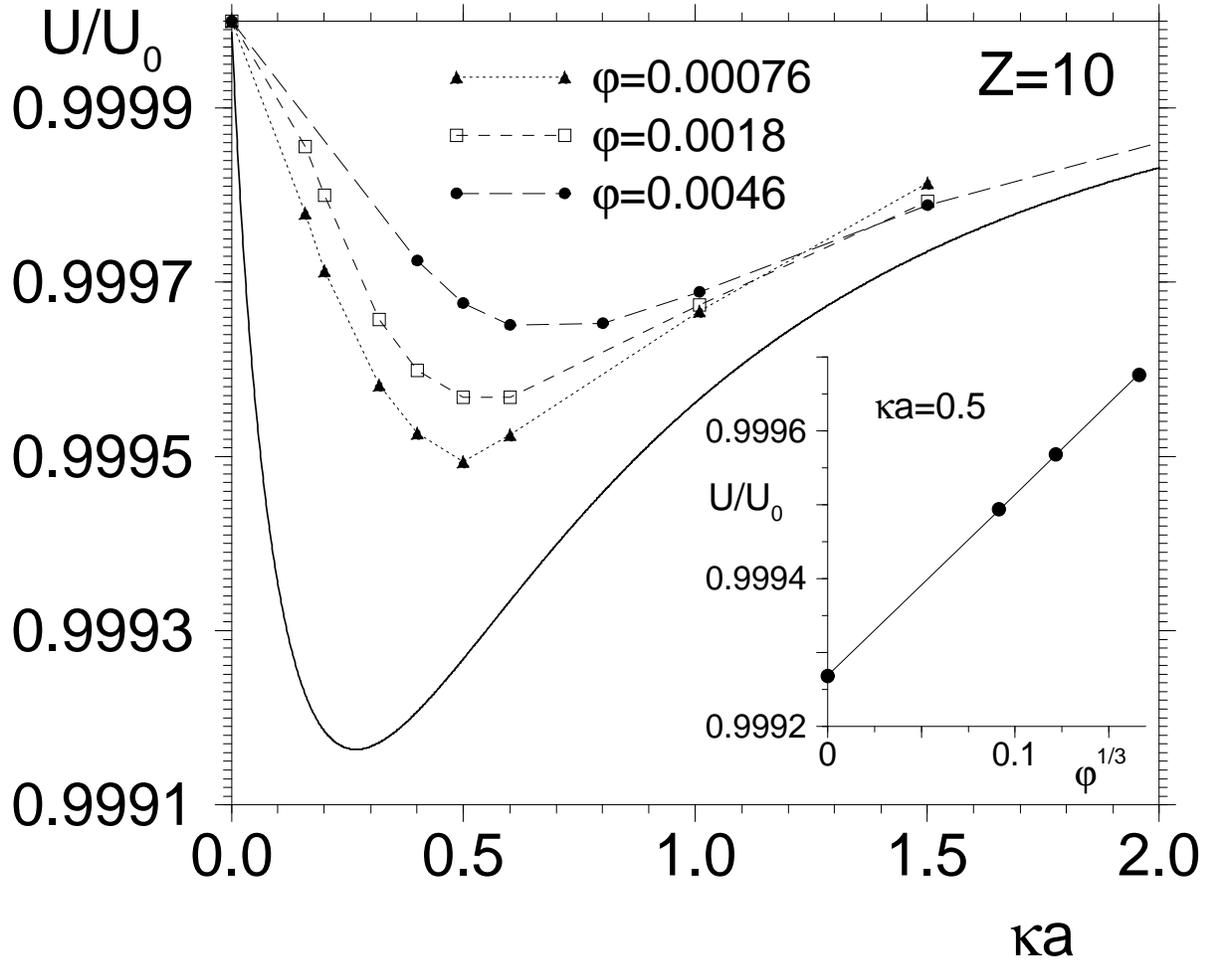,width=16cm,height=13cm}
\vspace{4mm}

\caption[fig2]{$U/U_0$ for the volume fractions $\varphi=$~0.0046, 0.0018, and 0.00076 
               as a function of $\kappa a$. The charge of the macroion is set to 
               $Z=10$. The solid line is the result from Booth's theory. The inset
               shows $U/U_0$ as a function of $\varphi^{1/3}$ for $\kappa a=0.5$. The
	       solid line in the inset is the fit 
	       function $0.999269+0.0024556 \varphi^{1/3}$.}
\label{fig2}
\end{figure}

Fig.~\ref{fig2} shows the results for a surface charge $Z=10$ of
the macroion.  We demonstrate below that this value is small enough
for the approximation (\ref{eq28}) to hold. Firstly, we infer from
Fig.~\ref{fig2} that the relative reduction of the sedimentation velocity
due to the charges is only of the order of $10^{-4}$ at $Z=10$ in the
$\varphi$--range considered. $U/U_0$ exhibits a minimum which moves to
higher values of $\kappa a$ with increasing $\varphi$. The occurence
of such a minimum is reasonable since an increase of $\kappa a$ is
accompanied with two competing effects: On the one hand the electrostatic
potential $\hat{\Phi}$ becomes stronger due to an increasing salt
concentration but on the other hand it becomes also more short--ranged
because of a decreasing Debye length, and thus, it affects only the flow
nearby the macroion. The value of the minimum in $U/U_0$ decreases 
with decreasing $\varphi$ and seems to move towards the one of the 
Booth curve for $\varphi \to 0$. This also holds for the amplitude 
and the shape of $U/U_0$ for $\varphi \to 0$.
In order to give quantitative evidence that our calculation
would recover Booth's result we plot in the inset
of Fig.~\ref{fig2} $U/U_0$ as a function of $\varphi^{1/3}$ at 
$\kappa a = 0.5$, i.e.~around the position of the minimum.
The fit in this figure with a straight line indeed approaches the 
Booth result, i.e.~at $\varphi=0$. 

\begin{figure}[tb]
\psfig{file=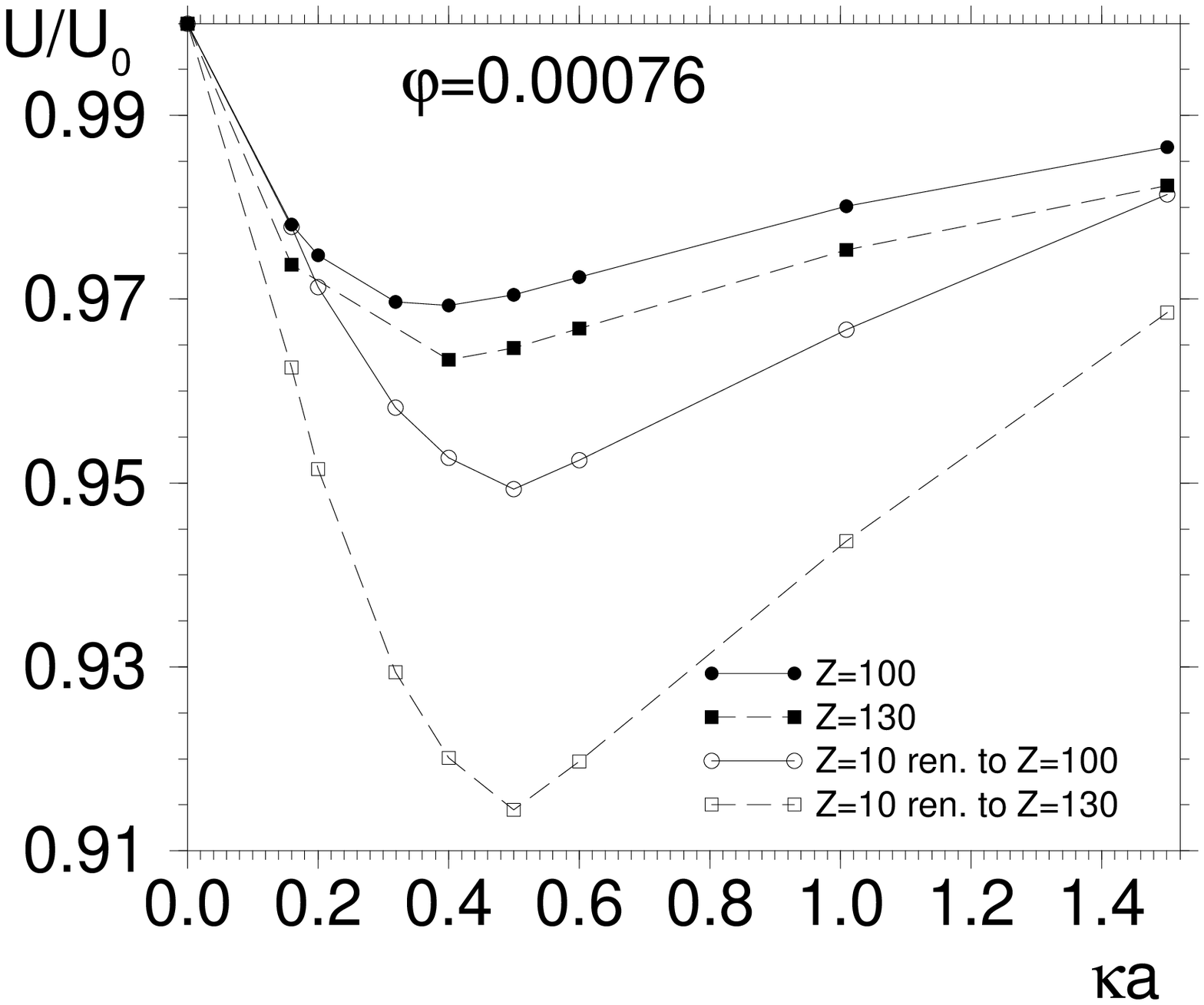,width=11cm,height=8cm}
\vspace{2mm}
\psfig{file=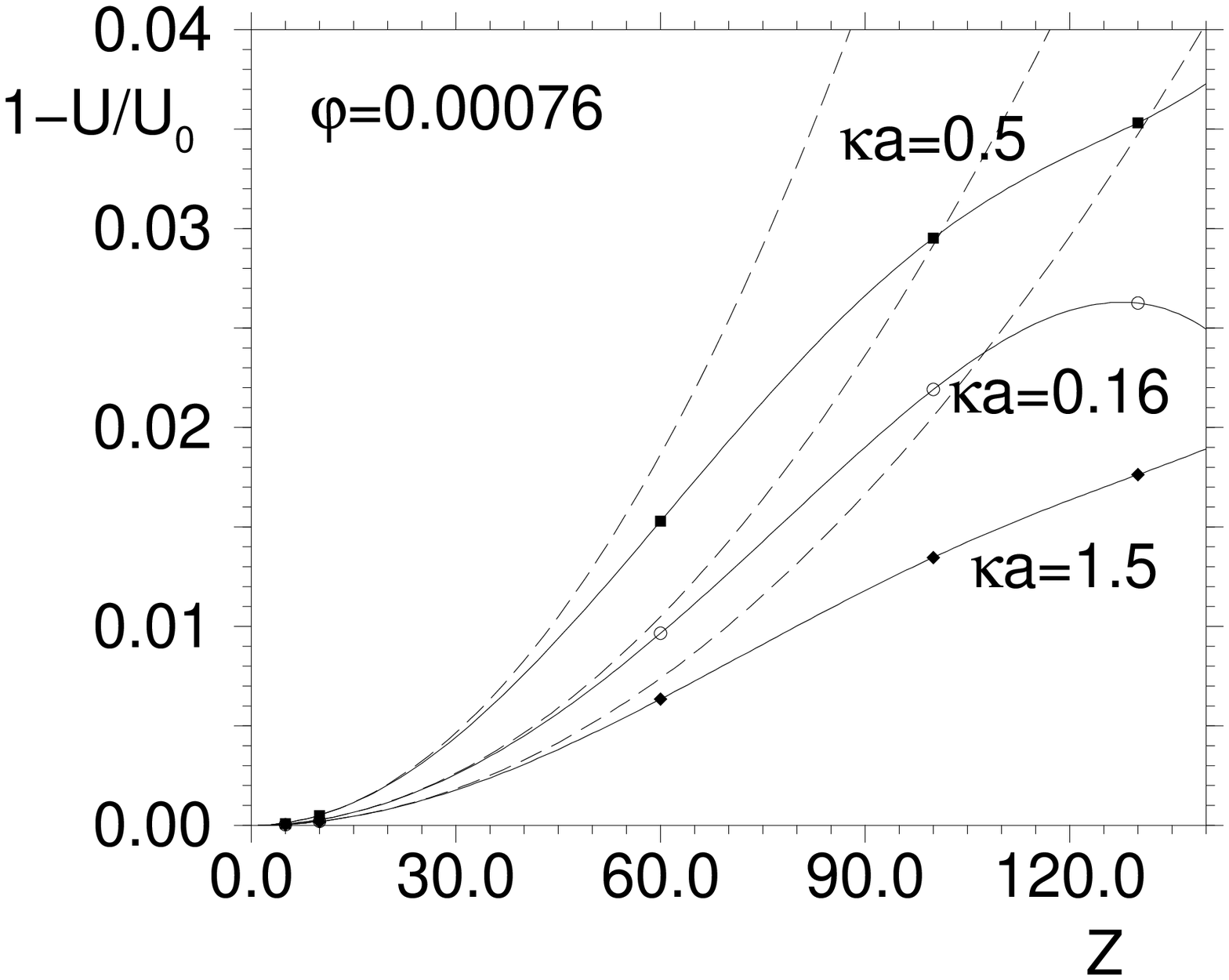,width=11cm,height=8cm}
\vspace{2mm}

\caption[fig3]{a) $U/U_0$ at $\varphi=0.00076$ as a function of $\kappa a$
                  for $Z=100$ and $Z=130$ (filled symbols). The open symbols
                  show data for $Z=10$ which are renormalized to $Z=100$
                  and $Z=130$ (see text).
               b) $1-U/U_0$ as a function of $Z$ for the indicated values of 
                  $\kappa a$. The solid lines show the following fit functions:
  $g(Z)=2.916 \cdot 10^{-6} Z^2 - 6.086 \cdot 10^{-11} Z^4 - 1.173 \cdot 10^{-15} Z^6$
  for $\kappa a = 0.16$,
  $g(Z)=5.182 \cdot 10^{-6} Z^2 - 2.810 \cdot 10^{-10} Z^4 + 5.801 \cdot 10^{-15} Z^6$
  for $\kappa a = 0.5$,
  $g(Z)=2.056 \cdot 10^{-6} Z^2 - 8.684 \cdot 10^{-11} Z^4 + 1.592 \cdot 10^{-15} Z^6$
  for $\kappa a = 1.5$. The dashed lines show only the term $\propto Z^2$ of the latter 
  three functions.}
\label{fig3}
\end{figure}

Up to now we have shown only the results for $U/U_0$ for a small charge
of the macroion $Z=10$. But it is of course interesting to check
up to which values of $Z$ the approximation (\ref{appbooth}) holds.
If Eq.~(\ref{appbooth}) would be exact, one could renormalize $U/U_0$
as a function of $\kappa a$ for a given charge $Z=Z_{{\rm old}}$ to
a new charge $Z=Z_{{\rm new}}$ by multiplying $1-U/U_0$ by $Z_{{\rm
new}}^2/Z_{{\rm old}}^2$. In this way we have renormalized our data
for $Z=10$ at $\varphi=0.00076$ to $Z=100$ and $Z=130$, and we compare
these data sets in Fig.~\ref{fig3}a with the corresponding simulation
results for the latter two values of $Z$.  We see that we have strong
corrections to the results as expected from Booth's theory, especially
around the minimum in $U/U_0$. Firstly the amplitude of the minimum
is underestimated by the renormalized curves and also the position of
the minimum is at a slightly larger value. To study the corrections to
Booth's theory more quantitatively we plot in Fig.~\ref{fig3}b $1-U/U_0$
as a function of $Z$ for $\kappa a=0.16, 0.5$, and $1.5$. The solid
lines in this figure are fits of the form $g(Z)=c_2 Z^2 + c_4 Z^4 +
c_6 Z^6$.  From the comparison of the different functions $g(Z)$
to the corresponding ones with only the leading term $\propto Z^2$
(dashed curves in Fig.~\ref{fig3}b) we can conclude that the corrections
to Booth's theory become important for $Z>50$.

\begin{figure}[tb]
\psfig{file=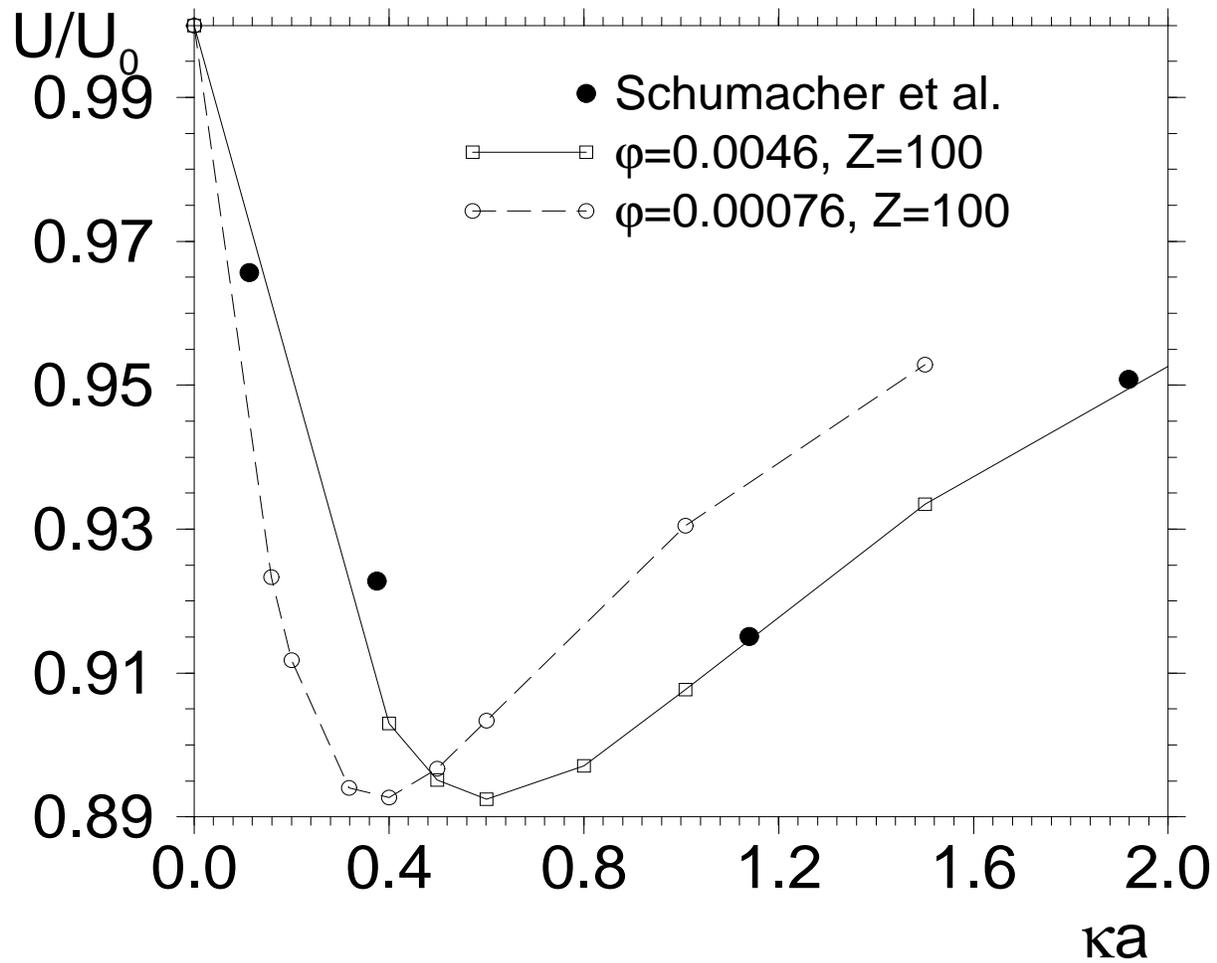,width=16cm,height=13cm}
\vspace{2mm}

\caption[fig4]{$U/U_0$ at $\varphi=0.00076$ and $\varphi=0.0046$ as a function of $\kappa a$
               for $Z=100$ in comparison to experimental data (closed circles).} 
\label{fig4}
\end{figure}

Finally we address the question whether our results are in agreement
with a dynamic light scattering experiment by Schumacher and van de
Ven~\cite{schumacher91}. They measured the diffusion constant $D$ for
a system of gold particles in distilled water.  Note that $D/D_0$ is
equal to $U/U_0$.  The radius of the gold particles was around 20~nm
and their volume fraction $2 \cdot 10^{-5}$~\%. Different data sets
were determined by changing the value of $\kappa a$ with different
salts.  We consider here the experimental data points measured with
sodium chloride which are shown in Fig.~\ref{fig4} in comparison
to our simulation data at $\varphi=0.0046$ and at $\varphi=0.00076$
for $Z=100$. It is interesting that the experimental data can be very
well described by the simulation curve for $\varphi=0.0046$ although the
experiment was done at a very small volume fraction of gold particles and
the curve for $\varphi=0.00076$ deviates strongly from the experimental
data. More systematic experiments, e.g.~for different volume fractions,
would be necessary to clarify this discrepancy.

\section{Conclusions}
We have developed a LBM for the simulation of the dynamics of suspensions
of charged colloidal particles. In this method a set of non--linear,
coupled electrokinetic equations is solved which consists of convective
diffusion equations for the ion densities $\rho_s$ ($s=+,-$),  the
linearized Navier--Stokes equations for the mass current $\vec{j}$,
and the  Poisson equation for the electrostatic potential $\hat{\Phi}$.
Furthermore, a neutral fluid characterized by the density $\rho_n$ can
be introduced in order to keep the viscosity in the fluid essentially
constant.  The propagation of $\rho_s$ and $\vec{j}$ is computed by means
of one--particle distribution functions $n_i^s$ for each species $s$.
But in contrast to the normal LBM $\rho_s$ and $\vec{j}$ are not simply
given as zeroth and first moments of the $n_i^s$'s, respectively.
This is due to the coupling of the ionic part of the fluid to the
gradient of $\hat{\Phi}$ which leads to an additional diffusive term
in the propagation of the ion densities and a body force term in the
propagation of the mass current $\vec{j}$. The electrostatic potential is
determined from the ion densities by means of a Poisson equation solver
for which we use a successive overrelaxation scheme. Our method is fast,
stable and easy to implement.  We have verified our numerical scheme by
means of a Chapman--Enskog expansion.
 
As an application we applied our method to determine the reduced
sedimentation velocity $U/U_0$ for an array of charged spheres on
a simple cubic lattice. We determined $U/U_0$ as a function of the
dimensionless parameter $\kappa a$ for different volume fractions of
macroions $\varphi$. We compared our results with an analytical formula
first derived by Booth which is valid in the limit of a weakly charged,
isolated macroion in an unbounded electrolyte. Booth's theory starts with
an expansion of $U/U_0$ in powers of the macroion charge $Z$ of which
the lowest non--vanishing order, $Z^2$ is taken into account. We gave
evidence that we recover Booth's result in the limit $\varphi \to 0$.
For about $Z>50$ corrections to the Booth theory become important.
Up to $Z=130$ our simulation data can be well described by an expansion
up to order $Z^6$. These numerical data could be used to test theories
that go beyond the Booth level. We mention that very recently also a
mode coupling theory with hydrodynamic interactions was shown to be in
agreement with Booth's theory \cite{kollmann00}. This theory is also
able to consider colloidal systems at finite volume fractions.

Our LBM for charged colloids is well suited to study their short--time dynamics
and the flow around macroions. It is not restricted to steady state problems
but one can also determine time dependent quantities like the velocity
autocorrelation function of a tagged macroion in a colloidal suspension.
Also particle shapes which are different from a spherical one can be
easily introduced in our LBM. Moreover, the introduction of walls is rather simple
which makes it possible to study charged colloidal particles in confined 
geometries.

Acknowledgments: We thank F.~Capuani for a critical reading of the manuscript.
The work of the FOM Institute is part of the scientific program
of FOM and is supported by the Nederlandse Organisatie voor Wetenschappelijk
Onderzoek (NWO). J. H. acknowledges financial support by the Deutsche
Forschungsgemeinschaft (Grant HO 2231/1-1).


\end{document}